\documentclass[a4paper,11pt]{article}
\pdfoutput=1 

\usepackage{jinstpub} 
\usepackage{lineno}

\newcommand{\typei}{Type $\rm{\,I\,}$ }
\newcommand{\typeii}{Type $\rm{I\hspace{-.01em}I}$ }

\title{Performance of RHICf detector during operation in 2017}

\collaboration{RHICf Collaboration}

\author[a,b]{O. Adriani,}
\author[a,b]{E. Berti,}
\author[b]{L. Bonechi,}
\author[a,b]{R. D’Alessandro,}
\author[c,d]{Y. Goto,}
\author[e]{B. Hong,}
\author[f,g]{Y. Itow,}
\author[h]{K. Kasahara,}
\author[e]{M. H. Kim,}
\author[f,1]{H. Menjo,\note{Corresponding author.}}
\author[c,d]{I. Nakagawa,}
\author[i]{T. Sako,}
\author[j]{N. Sakurai,}
\author[f]{K. Sato,}
\author[c,d]{R. Seidl,}
\author[k]{K. Tanida,}
\author[l]{S. Torii,}
\author[m,n,o]{and A. Tricomi}

\affiliation[a]{Department of Physics and Astronomy, University of Florence, Sesto Florentino (FI) I-50019, Italy}
\affiliation[b]{INFN Section of Florence, Sesto Fiorentino (FI) I-50019, Italy}
\affiliation[c]{RIKEN Nishina Center for Accelerator-Based Science, Wako, Saitama 351-0198, Japan}
\affiliation[d]{RIKEN BNL Research Center, Brookhaven National Laboratory, Upton, New York 11973-5000, USA}
\affiliation[e]{Korea University, Seoul 02841, Korea}
\affiliation[f]{Institute for Space-Earth Environmental Research, Nagoya University, Nagoya, Aichi 464-8602, Japan}
\affiliation[g]{Kobayashi-Maskawa Institute for the Origin of Particles and the Universe, Nagoya University, Nagoya, Aichi 464-8602, Japan}
\affiliation[h]{Shibaura Institute of Technology, Minuma-ku, Saitama 337-8570, Japan}
\affiliation[i]{Institute for Cosmic Ray Research, University of Tokyo, Kashiwa, Chiba 277-8582, Japan}
\affiliation[j]{Tokushima University, Tokushima, Tokushima 770-8051, Japan}
\affiliation[k]{Advanced Science Research Center, Japan Atomic Energy Agency, Tokai-mura, Ibaraki 319-1195, Japan}
\affiliation[l]{RISE, Waseda University, Shinjuku, Tokyo 162-0044, Japan}
\affiliation[m]{Department of Physics and Astronomy, University of Catania, Catania I-95123, Italy}
\affiliation[n]{INFN Section of Catania, Catania I-95123, Italy}
\affiliation[o]{CSFNSM, Catania I-95123, Italy}

\emailAdd{menjo@isee.nagoya-u.ac.jp}

\abstract{In the RHIC forward (RHICf) experiment, an operation with {\it pp} collisions was performed at $\sqrt{s}\,=\,$510 GeV from 24--27 June 2017. The performances, energy and position resolutions, trigger efficiency, stability, and background during the operation, have been studied using data and simulations, which revealed that the requirements for production cross-section and transverse single-spin asymmetry measurements of very forward photons, $\pi^0$s, and neutrons were satisfied. In this paper, we describe the details of these studies.}


\begin{document}
\maketitle
\flushbottom

\section{Introduction}

The relativistic heavy-ion collider forward (RHICf) experiment~\cite{rhicfloi} measured energetic neutral particles, photons, neutrons, and $\pi^0$s that were produced in a very forward region of polarized proton--proton collisions with a compact calorimeter detector installed at RHIC. The measurement was performed for two purposes and physics motivations: the first objective aimed to measure the differential production cross-section of neutral particles for testing and tuning the hadronic interaction models used for cosmic-ray air shower simulations. The second objective aimed to measure the transverse single-spin asymmetry of the particle production in the very forward region. 

A precise understanding of hadronic interactions is vital for enabling very high-energy cosmic-ray observations using the extensive air shower technique. In context, the mass composition of primary cosmic rays can be estimated using certain shower parameters that are observed by ground detectors, such as the depth of maximum shower development $X_{\mathrm{MAX}}$~\cite{PAO_Composition_2014_1,TA_Composition_2018}. 
However, the understanding of the results strongly relies on the selection of hadronic interaction models in air shower simulations. This is caused by the inadequate calibration data, especially the forward particle production relevant to air shower development. 
The LHCf experiment measured very forward photons, $\pi^0$s, and neutrons at LHC~\cite{LHCf_Photon_0.9TeVpp,LHCf_Photon_7TeVpp,LHCf_Photon_13TeVpp,LHCf_Pi0_Run1,LHCf_Neutron_7TeVpp,LHCf_13TeVpp_Neutron,LHCf_13TeVpp_Neutron2}. Likewise, RHICf performed a similar measurement at RHIC with a relatively lower center-of-mass collision energy of 510 GeV than that of 0.9-13 TeV conducted at LHC. 
In comparison to the LHCf results, the energy scaling of forward particle production can be tested, and it will contribute toward improving the predictions of models with intermediate and even higher collision energies. 

As discovered in the middle and forward pseudorapidity regions, large transverse single-spin asymmetry ($A_{N}$) of $\pi^0$ demonstrated an evident dependency on the transverse momentum $p_{\mathrm{T}}$ and the Feynman-x $x_{\mathrm{F}}$ of $\pi^0$~\cite{STARpi0,RHENIXpi0} with polarized beams of RHIC. 
The RHICf experiment extends the measurement to the very forward region corresponding to a low $p_{\mathrm{T}}$ of $<1$ GeV/c and large $x_{\mathrm{F}}$. 
Moreover, the RHICf has already reported the discovery of unexpectedly large transverse single-spin asymmetry in forward $\pi^0$ production~\cite{RHICf_pi0_An}.

An operation involving proton--proton collisions with the center-of-mass collision energy of $\sqrt{s}$ = 510 GeV was successfully completed in June 2017. In this paper, we present the performance of the RHICf detector during this operation. Therefore, a performance similar to that of the LHCf detector is required to achieve these goals: high detection efficiency with $x_{\mathrm{F}} > 0.1$, adequate energy resolution for photons ($<$ 5\%), and impact position resolution ($<$ 0.2 mm). 
Furthermore, we analyzed the obtained data and reported the background level and stability of the energy scale. 
The detector, location, and data acquisition system of the experiment are explained in Section~\ref{sec:setup}. After introducing the analysis method in Section~\ref{sec:analysis}, the operation conditions and detector performances are discussed in Sections ~\ref{sec:operation} and \ref{sec:performance}, respectively. Lastly, the results are summarized in Section~\ref{sec:summary}.  

\section{Experimental Setup}
\label{sec:setup}

\begin{figure}[tb]
    \centering
    \includegraphics[width=0.9 \textwidth]{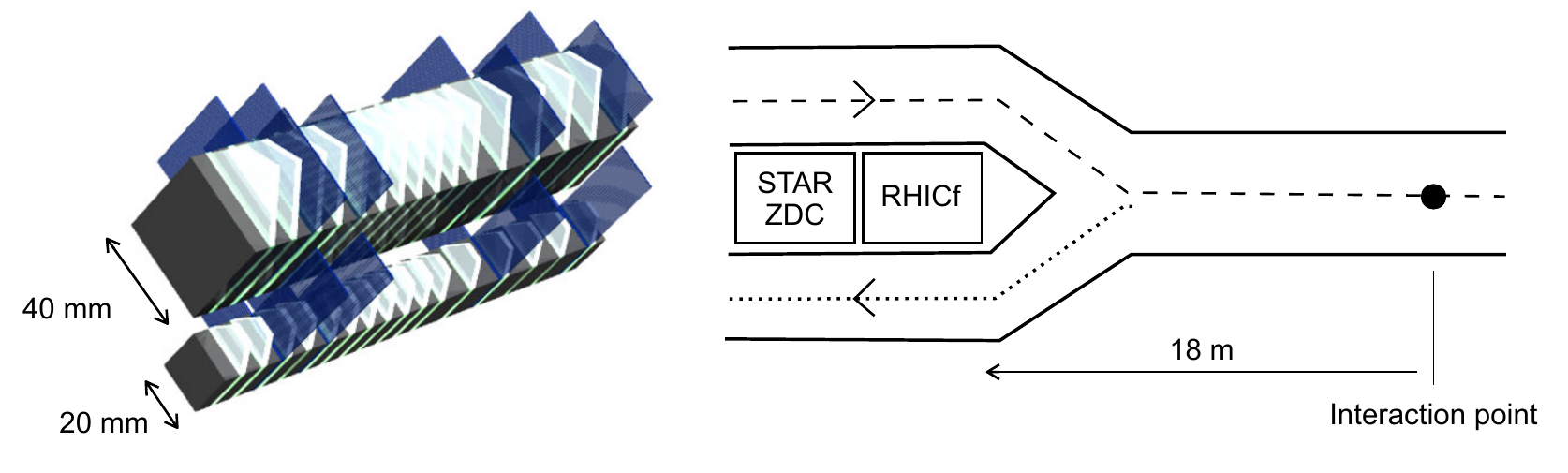}
    \caption{ Schematic (left) and location (right) of RHICf detector. }
    \label{fig:detector}
\end{figure}

\subsection{Detector and location}
The RHICf detector comprises two compact sampling and positioning calorimeters, which is actually the former LHCf-Arm1 detector \cite{LHCf_TDR}. Each calorimeter tower has dimensions transverse to the beam direction of 20\,mm\,$\times$\,20\,mm or 40\,mm\,$\times$\,40\,mm, and a longitudinal size of 220\,mm composed of 44\,radiation lengths of tungsten, as shown in Figure ~\ref{fig:detector}. In addition, 16 GSO scintillator plates with 1 mm thickness were inserted at every 2 or 4 radiation lengths for recording the longitudinal shower sampling, and four position-sensitive layers containing an X--Y hodoscope with 1 mm $\times$ 20 or 40 mm GSO bars \cite{LHCf_GSObar} were inserted at 6, 10, 30, and 42 radiation lengths for measurements of the lateral shower development.
Hereafter, the small and large calorimeter towers are referred to as TS and TL, respectively. 
The light from each scintillator plate and bar was transferred through a light guide and measured using a photomultiplier tube (PMT, Hamamatsu R7400U) and a multi-anode PMT (HAMAMATSU H7456), respectively. 
In addition, a 3\,mm thick scintillator termed as Front Counter was installed at the front of the calorimeter detector for the potential identification of background charged hadrons (protons and charged pions). 

The RHICf detector was installed at 18\,m west from the Solenoidal Tracker At RHIC (STAR) interaction point~(IP), wherein two beam pipes were situated with a 10\,cm gap to allow the insertion of the RHICf detector. In this setup, the detector can observe the very forward region of collisions, including zero degrees. 
Moreover, there are dipole magnets between the IP and the RHICf detector. The charged particles produced during proton--proton collisions were swept out by the magnetic field, so the RHICf detectors could measure only the neutral particles, photons, and neutrons.
The stated location is generally occupied by the STAR Zero-Degree Calorimeter (ZDC) ~\cite{star_ZDC}. During the RHICf operation, the ZDC was receded, and the RHICf detector was installed immediately in front of the ZDC, as shown in Figure ~\ref{fig:detector}.

\begin{figure}[tb]
    \centering
    \includegraphics[width=0.8 \textwidth]{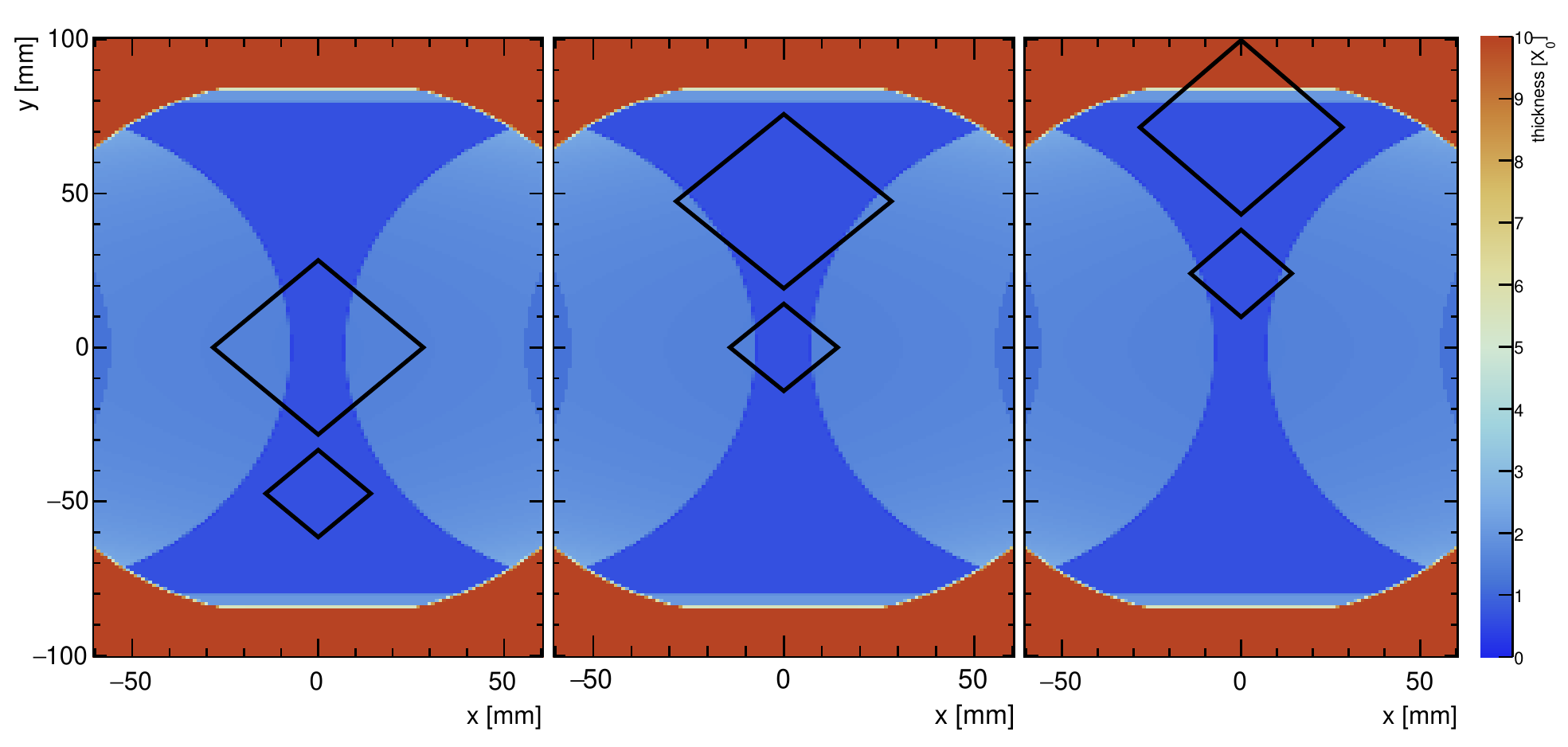}
    \caption{Thickness of beam-line materials between the IP and RHICf detector in a unit of radiation length $X_{0}$. 
    Ideal beam center position represents origin of coordinate system. Three panels signify the same map with various references of RHICf calorimeter towers at BOTTOM (left), MIDDLE (center), and TOP (right) positions identified by two black diamond shapes. }
    \label{fig:beampipe}
\end{figure}

Figure~\ref{fig:beampipe} illustrates the effective thickness of the beam-line structure between the IP and the detector as functions of the vertical and horizontal positions on the detector plain.
The acceptance was restricted by the structure to connect the beam pipe of 10\,cm radius to two small beam pipes located in front of the detector, which corresponds to the red-colored region in Figure ~\ref{fig:beampipe}. 
The detector was lifted by a manipulator and could be moved vertically. The operation was performed in three positions: BOTTOM, MIDDLE, and TOP, representing the diamond shapes in Figure ~\ref{fig:beampipe} to cover the full acceptance at the location. The pseudorapidity coverage was $\eta\,>\,6.05$.

\subsection{Trigger and data acquisition system}
Trigger signals were generated with a logic based on the energy deposition on the scintillator plates, which was implemented on a system with discriminator modules and a field-programmable gate array (FPGA) board. In the logic, three trigger modes were implemented to efficiently obtain interesting events for conducting the present physics analyses with a limited data acquisition (DAQ) speed as follows: 

\begin{itemize}
\item {\bf Shower trigger} was implemented for detecting any electromagnetic (EM) and hadronic showers induced by photons and neutrons. This trigger was issued in case any of the three successive layers have hits of the discriminator, for which the threshold was set to approximately 45 MeV.
The fourth layers in both the towers were not included in this logic, and they were assumed to have hits always owing to the higher discriminator threshold of the high-EM trigger described in the following. 
Additionally, the \item {\bf ${\bf \pi^{0}}$ trigger} was specialized to detect photon pairs from the $\pi^{0}$ decays, corresponding to \typei $\pi^0$ events defined in Section~\ref{sec:pi0}. This trigger was issued in case EM showers were simultaneously detected in both the calorimeters. Moreover, the trigger generation by hadronic showers was minimized by considering the shower trigger condition in the shallow layers from the first to seventh layers. 
Similarly, the \item {\bf High-EM trigger} was implemented to increase the statistics of high-energy photon events ($>$ 100 GeV), which were relatively rare owing to the soft photon production spectrum. The trigger was issued when the fourth layer in either of the calorimeter exhibited an energy deposit greater than ~500 MeV. In this trigger sample, \typeii $\pi^0$ events were enriched as well, in which a photon pair from a $\pi^0$ decay hit one calorimeter and possessed a total photon energy > 100 GeV.
\end{itemize}

\begin{figure}[tb]
\begin{center}
\includegraphics[width=350pt]{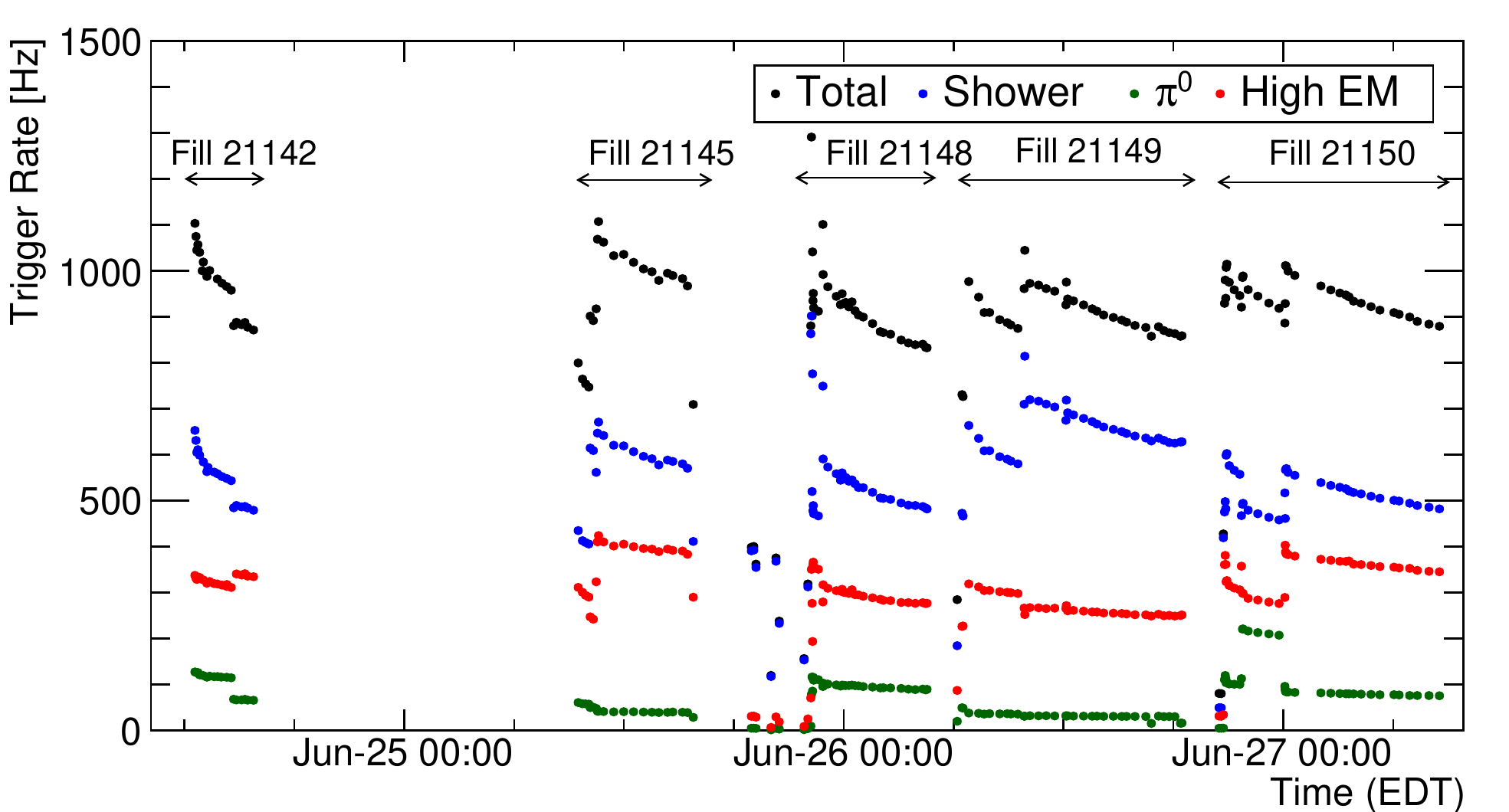}
\end{center}
\caption{Trigger rates of RHICf operation in 2017; upper arrows indicate physical operation periods.}
\label{fig_trgrate}
\end{figure}

These trigger signals were mixed after pre-scaling down by factors of 8--30 for shower triggers and 1--4 for high-EM triggers. The rate of $\pi^{0}$ triggers was the lowest among the three trigger modes, and its pre-scaling factor was set to 1 in all the cases, i.e., no pre-scaling was conducted. The event rates of these triggers strongly relied on the operating conditions, such as the vertical detector position. The raw rate of shower triggers was 6--30 kHz, which was scaled down to approximately 1 kHz after pre-scaling.
During the operation, these pre-scaling factors were occasionally optimized to maintain the final trigger rate at approximately 1 kHz, which corresponds to the DAQ condition with approximately 50\% of the DAQ live fraction. 
As can be observed from Figure~\ref{fig_trgrate} illustrating the total trigger rate along with the three trigger rates, the rates decreased during an operation period owing to the reduced beam intensity, whereas the rate suddenly increased at certain moments owing to re-optimisations of the pre-scaling factors.

The DAQ system was implemented using a LHCf system. The output data of analog-to-digital converters and other electronics were processed via a VME bus and stored on a local disk. The run control and front-end software were implemented based on the MIDAS package~\cite{MIDAS}. In particular, the read-out time for one event data was approximately 500~$\mu$s, and the maximum data acquisition rate was 2 kHz. 

The RHICf final trigger signals were transmitted to a central trigger system of the STAR experiment and triggered the STAR DAQ system to perform a common operation with STAR. The RHICf data were transmitted to the STAR DAQ system via a socket connection and stored in the STAR disk system along with the data acquired from the other STAR detectors.

\section{Event reconstruction}
\label{sec:analysis}

\begin{figure}[tb]
    \centering
    \includegraphics[width=0.8 \textwidth]{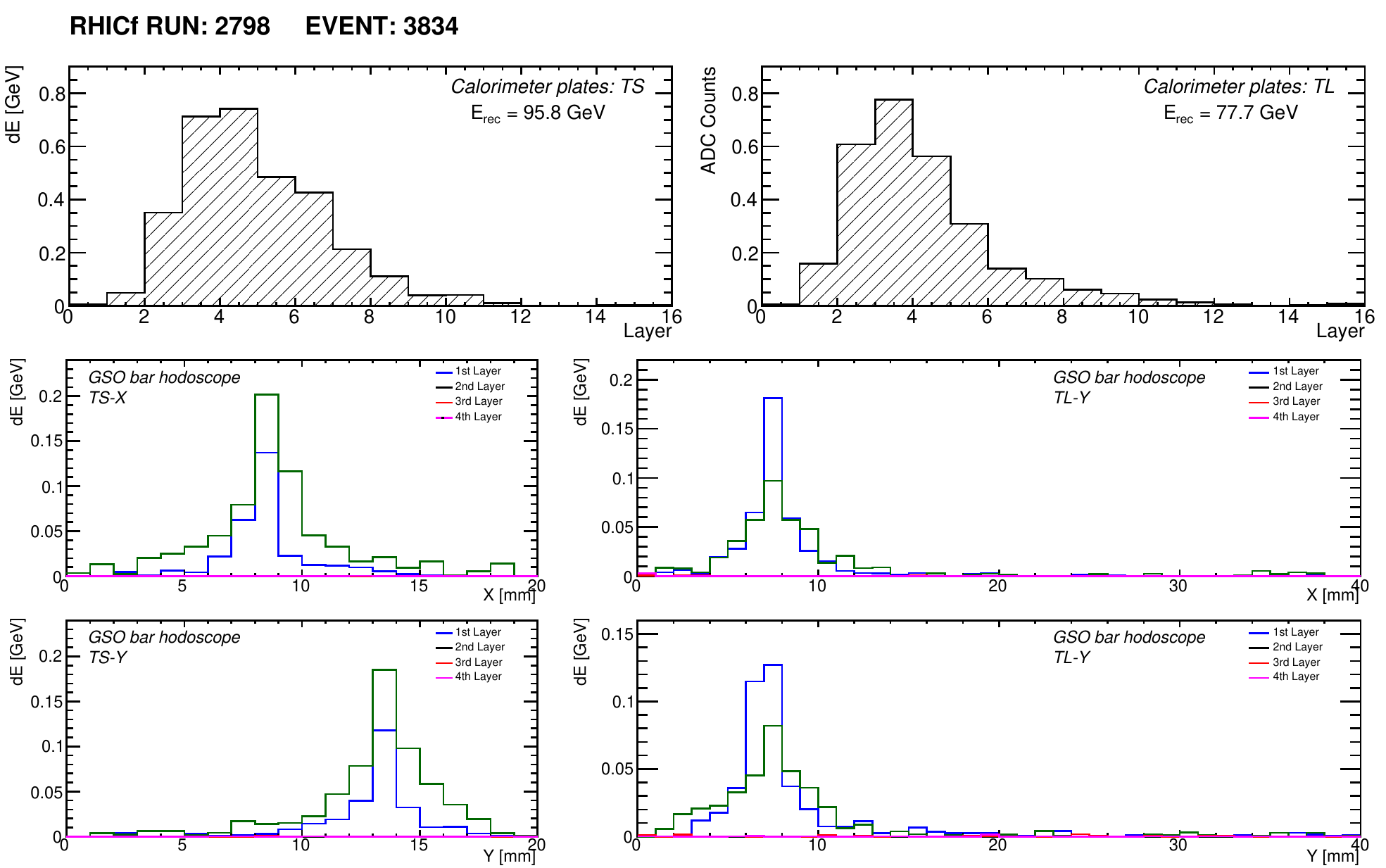}
    \caption{Example of an event measured by the RHICf detector. The event records a single photon incident in each calorimeter tower, which is a candidate for $\pi^0$. The top panels represent the longitudinal shower development measured by the scintillator layers of TS (left) and TL (right). The estimated photon energies in TS and TL were 95.8 and 77.7 GeV, respectively. The middle and bottom panels depict the lateral development measured by four layers of GSO bar XY-hodoscopes. Blue, green, red, and magenta lines indicate profiles at various layers: 1st (6 $X_{0}$), 2nd (10 $X_{0}$), 3rd (30 $X_{0}$), and 4th (42 $X_{0}$), respectively.}.
    \label{fig:eventview}
\end{figure}

In this section, we briefly introduce the event reconstruction method used in the subsequent analyses, which basically follows that of the LHCf analysis~\cite{LHCf_Photon_13TeVpp}. There are three main steps in the event reconstruction: position and energy reconstruction and particle identification (PID). Figure \ref{fig:eventview} presents an example of an event obtained during this operation, which is a Type $I$ $\pi^0$ event discussed in Section ~\ref{sec:pi0}.

\begin{figure}[tb]
    \centering
    \includegraphics[width=1.0 \textwidth]{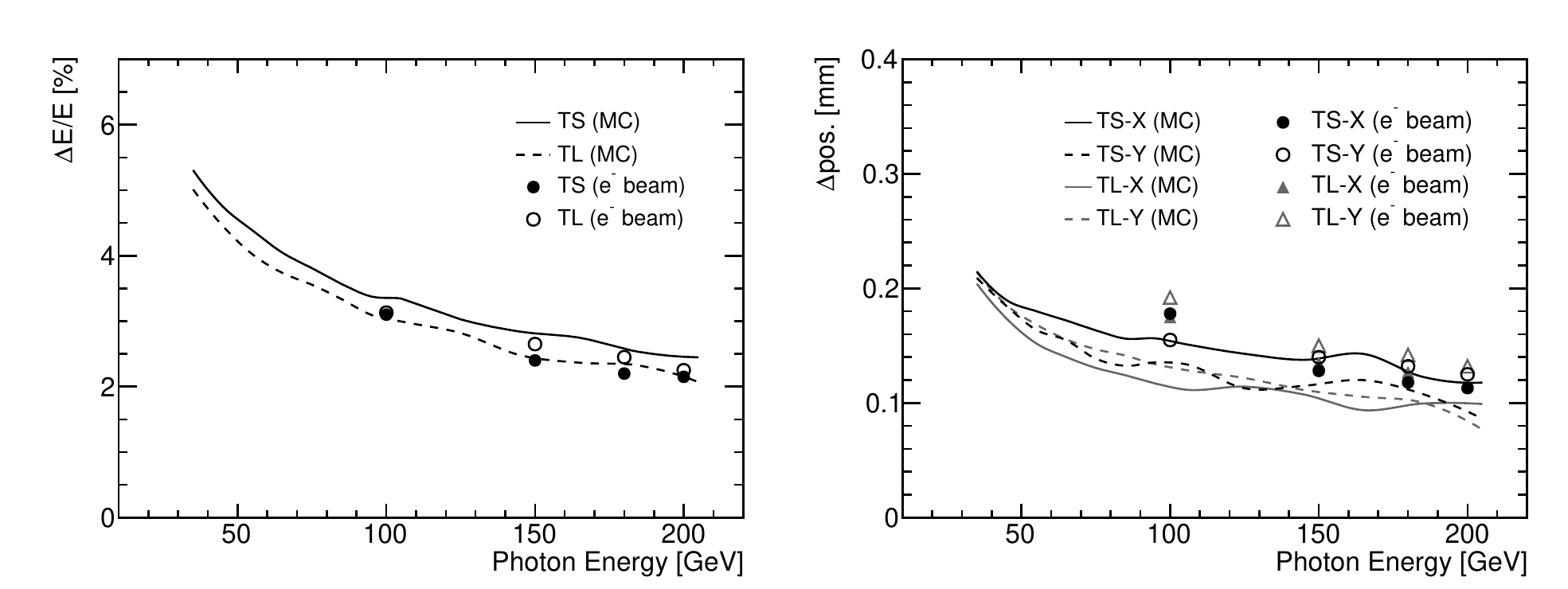}
    \caption{Energy (left) and position (right) resolutions for photons. Lines represent resolution estimated from full-detector simulation; markers represent resolutions obtained using electron beams at CERN SPS~\cite{LHCf_Photon_Performance_Makino} }.
    \label{fig:resolusion}
\end{figure}

The impact positions of the showers were determined from the fitting result of the lateral distribution measured using the four GSO bar hodoscope layers with the following function $f(x)$ motivated by the Lorentzian function:    
\begin{equation}
    f(x) = \frac{C}{s} [ \frac{\sigma_s\,a}{((x-x_{0})+\sigma_s)^{3/2}} + \frac{\sigma_w\,(1-a)}{((x-x_{0})+\sigma_w)^{3/2}} ] + C_0,  
\end{equation}
where $x_0$ corresponds to the impact position of the shower, $\sigma_s$ and $\sigma_w$ are width parameters describing sharp and wide distributions, respectively; $C$ and $a$ are normalization and fraction parameters of the two components, respectively, and $C_0$ is a constant term describing common noise. The position resolution for the photons estimated using a detector simulation was better than 0.2 mm, as shown in Figure \ref{fig:resolusion} (right). Moreover, the position reconstruction performance was measured using an electron beam with energies of 100--200 GeV~\cite{LHCf_Photon_Performance_Makino} (markers in Figure~\ref{fig:resolusion}), whereas the beam test results indicate slightly inferior simulation results owing to the uncertainty of the gain calibration for each bar.

The energy of an incident photon was estimated from the energy deposition on the scintillator plates as well as the reconstructed position. First, the summation of the energy deposition from the 2nd to the 12th layers, $S$, was calculated as: 
\begin{equation}
    S\,=\, \sum^{10}_{i=2} dE_{i} Y_{i}(x,y) L_{i}(x,y) S_{i}, 
    \label{eq:sumdE}
\end{equation}
where $dE_{i}$ is the measured energy deposition on the $i$-th scintillator plate, $Y_{i}(x,y)$ and $L_{i}(x,y)$ are the position-dependent correction factors of light yield on the $i$-th scintillator layer and leakage of shower particles from the calorimeter tower, respectively, and $S_{i}$ is a weight factor for correcting the various sampling steps (1 or 2). The factors $Y_{i}(x,y)$ and $L_{i}(x,y)$ were obtained from the pre-calibration of each scintillator plate and a detector simulation. The photon energy $E_{\gamma}$ was estimated from $S$ using a 1st polynomial function $F$ defined by the simulation as
\begin{equation}
E_{\gamma} = \alpha F(S),
\label{eq:energy}
\end{equation}
where $\alpha$ is an energy re-scaling parameter obtained from the $\pi^0$ analysis discussed in Section ~\ref{sec:pi0}. The neutron energy was estimated using a similar method by increasing the number of layers required for the summation of energy deposits.  
Consequently, the energy resolution estimated using the simulation accounting for actual electrical noise during the operation was 2--5\% for photons (Figure ~\ref{fig:resolusion} (left)) and 40\% for neutrons. In Figure ~\ref{fig:resolusion}, the results measured using electron beams at the CERN SPS are included as references~\cite{LHCf_Photon_Performance_Makino}.

\begin{figure}[tb]
    \centering
    \includegraphics[width=0.5 \textwidth]{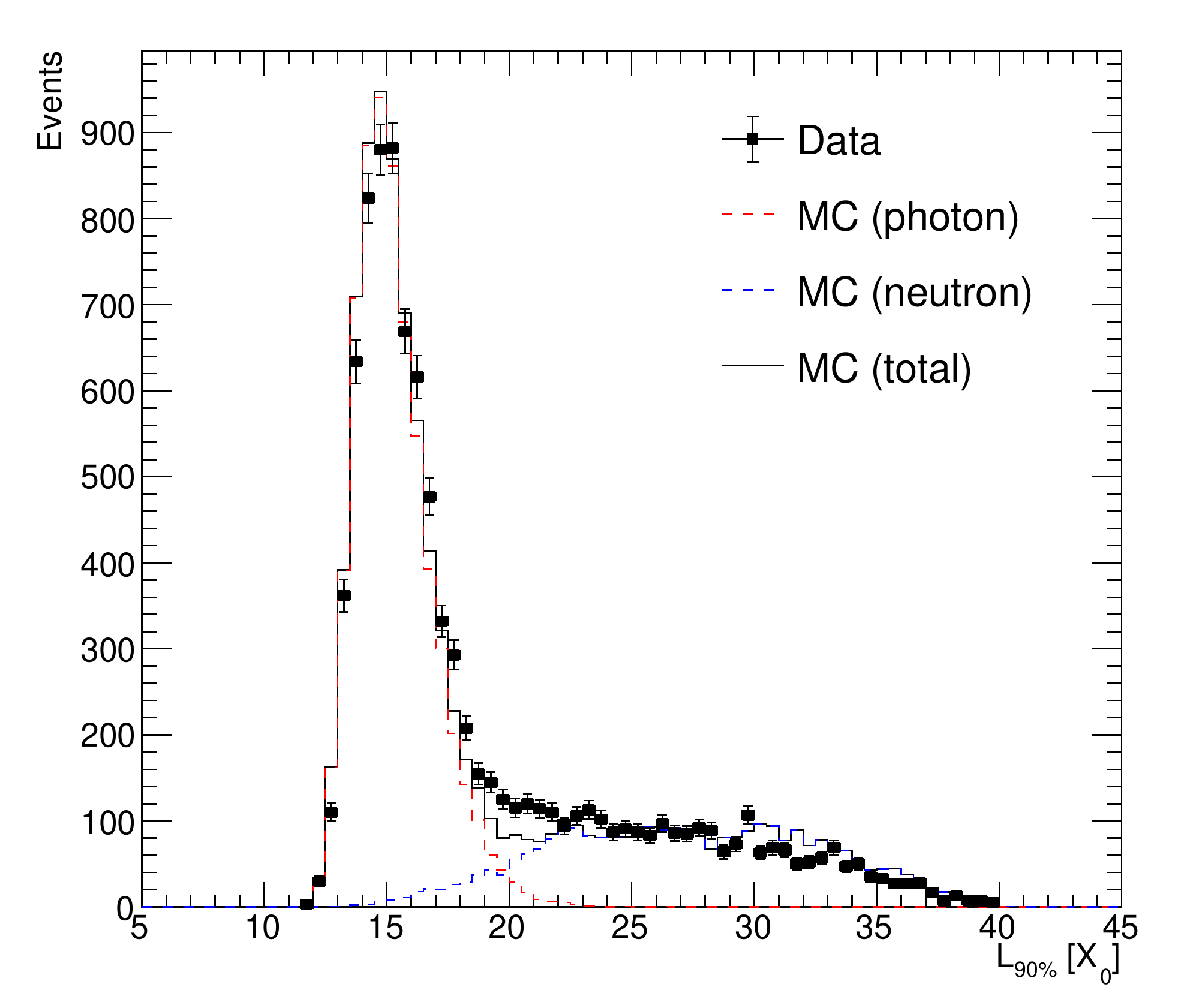}
    \caption{$L_{90\%}$ distribution of events in TL with reconstructed energies within 80-100 GeV. Template distributions normalized by the template-fit result for photons and neutrons are represented by red and blue dashed lines, respectively. Total number of template distributions is shown with black line.}
    \label{fig:l90}
\end{figure}

Particle identification was performed using different longitudinal developments in the calorimeter between electromagnetic and hadronic showers induced by incident photons and neutrons, respectively.
The photon candidate events were selected using a PID estimator $L_{90\%}$ that was defined in units of radiation length ($X_0$) as the longitudinal depth at which the energy deposit integral measured by the sampling layers reached 90\% of the total energy deposition. Figure~\ref{fig:l90} illustrates the $L_{90\%}$ distribution of the data with reconstructed energies between 80 and 100 GeV. The photon selection criterion was $L_{90\%}\, <\, L_{90\%,thr}$, where $L_{90\%,thr}$ reflects an $L_{90\%}$ threshold, thereby maintaining a 90\% survival efficiency of photons. Moreover, the inefficiency of photons and contamination of hadrons were estimated using a template-fit method of the $L_{90\%}$ distribution onto distributions of pure photon and neutron MC samples.

\section{Operation condition}
\label{sec:operation}

\subsection{Beam condition}
The RHICf operation was performed from 24th until 27th June, 2017 under a special low-luminosity condition. The physics data were obtained during five beam fills---referred to as RHIC Fill 21142, 21145, 21148, 21149, and 21150. In all the fills, 111 bunches of protons in 120 slots were accelerated to 255 GeV/$c$ in both clockwise and counterclockwise rotating beams, referred to as blue and yellow beams, respectively. Among them, 102 bunches collided at the IP. The minimum bunch spacing was 106 ns. 
The proton beams were radially polarized during RHICf operation, whereas the usual polarization is vertical. The direction of the beam polarization was rotated by $90^{\circ}$ using spin rotator magnets for radial polarization to maximize the sensitivity of the single spin asymmetry measurements. The polarization ranged from 0.53--0.59 with a systematic uncertainty of less than 0.02. 

The luminosity in the RHICf operation was set to approximately $~10^{31}$ $\mathrm{cm^{-2}s^{-1}}$, which is an order of magnitude lower than the nominal value of RHIC at {\it pp} collisions required to maintain a low collision pileup rate of $~0.05$ collisions per bunch crossing. This was realized with high $\beta^{*}$ optics of 8 m. In particular, a large $\beta^{*}$ is preferable to reduce the beam angular divergence and reduce the event-by-event fluctuation of the beam-center position on the detector plane. Furthermore, the temporal variation of the luminosity was monitored using the coincidence rate of both STAR-ZDC signals. The conversion factor from the ZDC signal rate to the absolute luminosity was determined from nine beam scans, i.e., Vernier scans, performed during the operation. The delivered integrated luminosity $L$ was 896 nb$^{-1}$ in total. 

The beam conditions for each fill are summarized in Table~\ref{tab:runtable}, as well as the operating conditions and statistics of the obtained data. 

\begin{table}[tb]
    \centering
    \caption{Summary of beam and operation conditions.}
    \begin{tabular}{c|c c c c c c}
        RHIC Fill & 21142 & 21145 & 21148 & 21149 & 21150 & Total \\
        \hline
        Duration [hours] & 2.6 & 4.6 & 4.1 & 7.5 & 8.1 & 26.9 \\
        Luminosity $\times10^{31}$ [$\mathrm{cm^{-2}s^{-1}}$] & 1.0-1.2 & 0.7-1.0 & 1.0-1.5 & 0.7-1.0 & 0.9-1.4 & \\
        Delivered $L$ [nb$^{-1}$] & 72 & 98 & 173 & 229 & 324 & 896  \\
        $\beta^{*}$ [m] & 8 & 8 & 8 & 8 & 8 & \\
        Vernier scans & - & 1 scan & 2 scans & 2 scans & 4 scans & \\
        Detector position & BOTTOM & BOTTOM & MIDDLE & TOP & MIDDLE & \\
        Statistics  &  &  &  &  & & \\
        \multicolumn{1}{l|}{\hspace{1.0cm} Shower trigger} & 5.0 M & 9.7 M & 7.7 M & 17.6 M & 14.7 M & 55 M\\
        \multicolumn{1}{l|}{\hspace{1.0cm} $\pi^{0}$ trigger} & 0.92 M & 0.71 M & 1.44 M & 0.88 M & 3.31 M & 7.3 M\\
        \multicolumn{1}{l|}{\hspace{1.0cm} High-EM trigger} & 3.1 M & 6.3 M & 4.4 M & 7.4 M & 9.8 M & 31 M\\ 
    \end{tabular}
    \label{tab:runtable}
\end{table}

\subsection{Beam center}

The average beam center position with respect to the projected position of zero-degree collisions or the beam direction on the detector plane is an important parameter in physical analyses to calculate the transverse momentum of incident particles from the measured energies and hit positions.

\begin{figure}[tb]
    \centering
    \includegraphics[width=0.5 \textwidth]{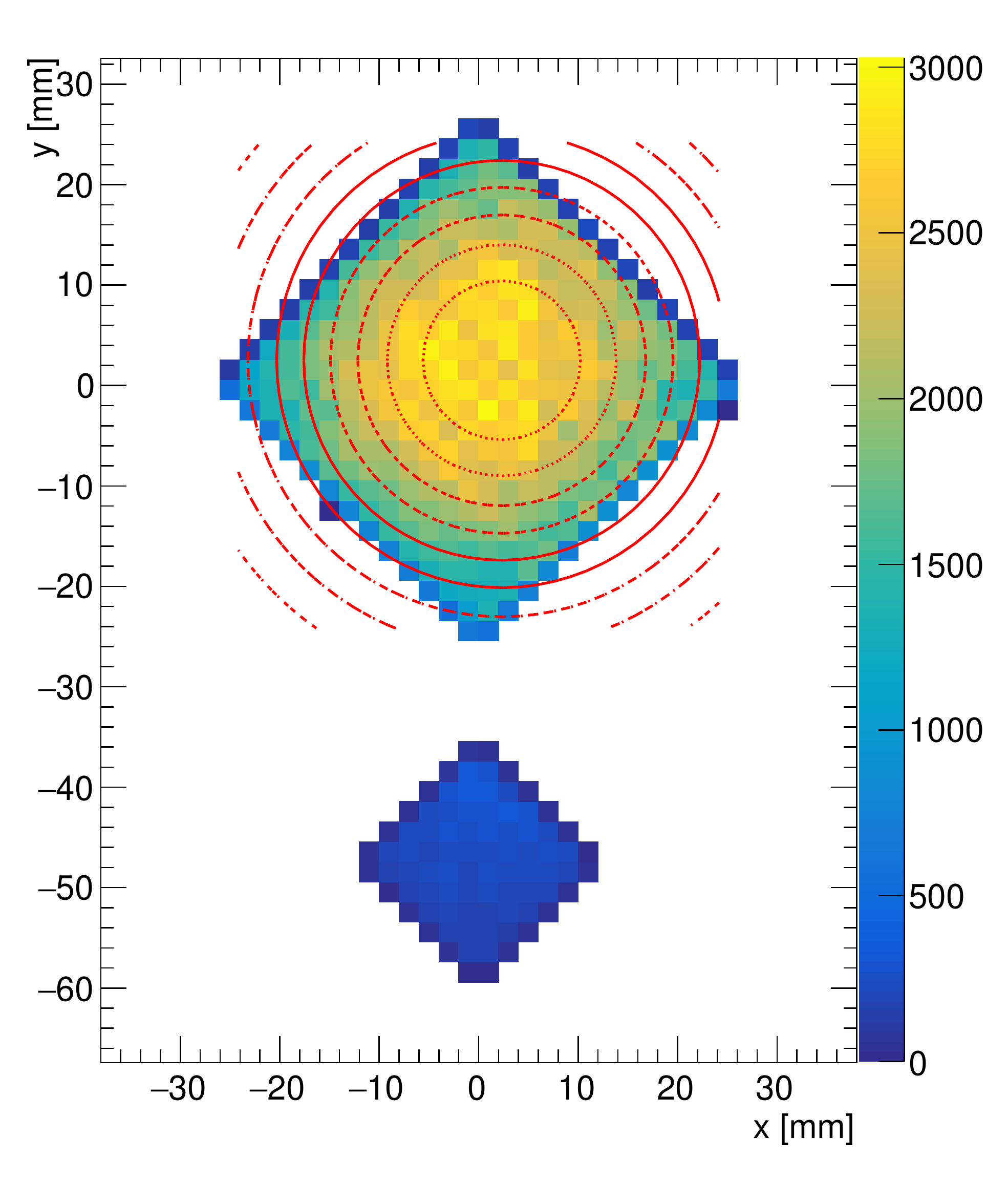}
    \caption{Hit position distribution of neutron events with reconstructed energies > 150 GeV. Color scale indicates the number of events recorded for each cell. Origin of coordinates was defined as center of TL tower. Contour represents the fitting result by a two dimensional Gaussian function.}
    \label{fig:neutronhitmap}
\end{figure}

In the BOTTOM position, the beam-center position was determined by fitting the measured hit-position distribution of neutrons with reconstructed energies > 150 GeV to a two-dimensional Gaussian function, as shown in Figure ~\ref{fig:neutronhitmap}. The obtained beam center position was $(x,y) = (2.3\pm0.1\,\mathrm{mm}, 2.5\pm0.1 \,\mathrm{mm})$ in the coordinate system presented in Figure ~\ref{fig:neutronhitmap}. This result was confirmed using another independent method that utilized the known spin asymmetry of forward neutron production~\cite{PhysRevD.88.032006}. 
This method was sensitive only to the vertical position, and the result was $y= 2.4\pm0.1$ mm. 

In the MIDDLE position, calculations of the beam-center position were performed using the same method as the BOTTOM position. However, the determination was more challenging owing to the small coverage by the TS near the beam center. Consequently, a difference of 1 mm was detected between the two methods, which was considered a systematic uncertainty in the physical analyses. In the TOP position, the beam center was not covered by the detector, and neither of the method worked. Therefore, the beam-center position was estimated from a mechanical measurement of the detector position, assuming the same beam center position relative to the beam pipe as that present in the other fills.

\subsection{Beam gas background}

\begin{figure}[tbp]
    \centering
    \includegraphics[width=0.45 \textwidth]{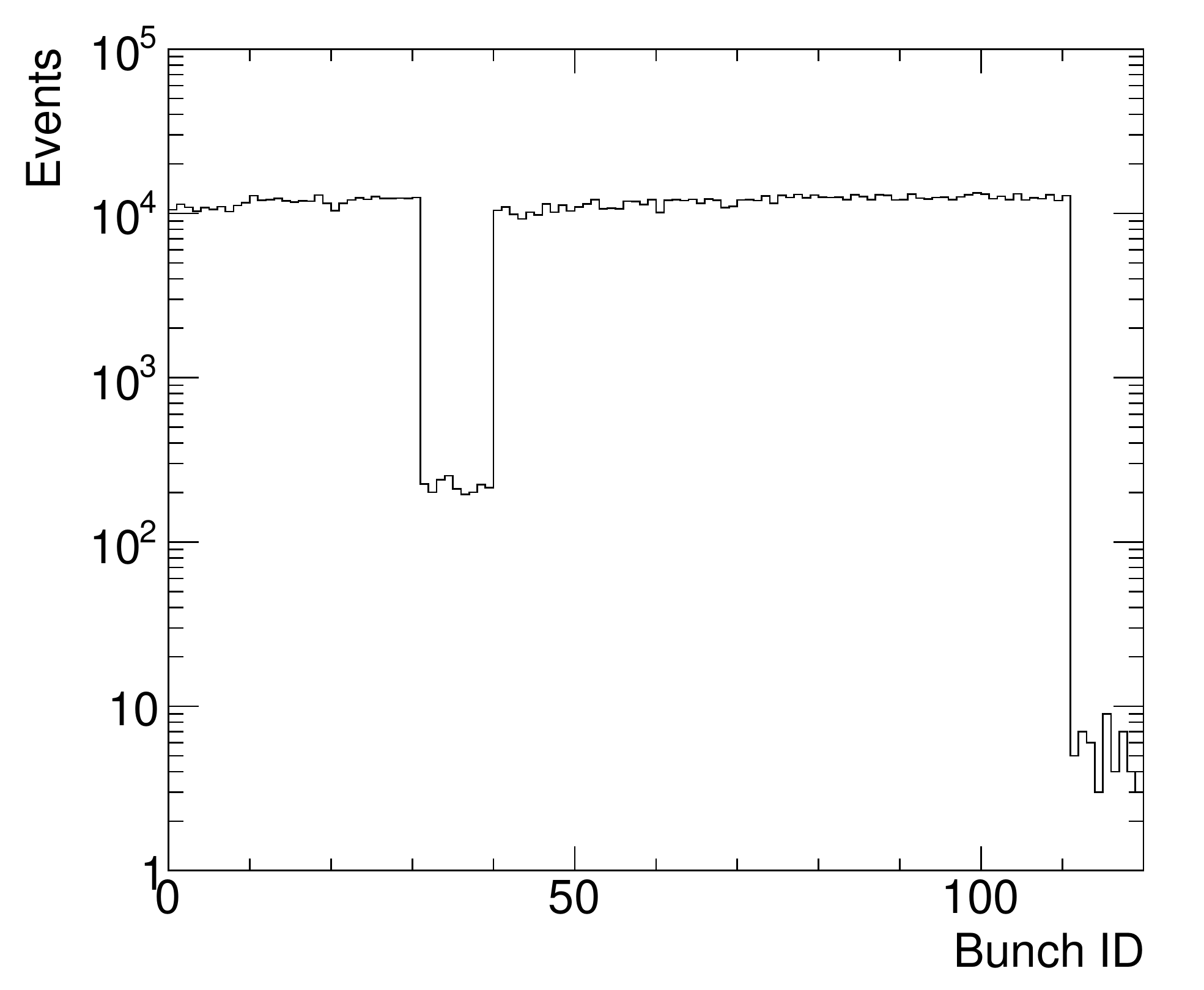}
    \includegraphics[width=0.45 \textwidth]{ 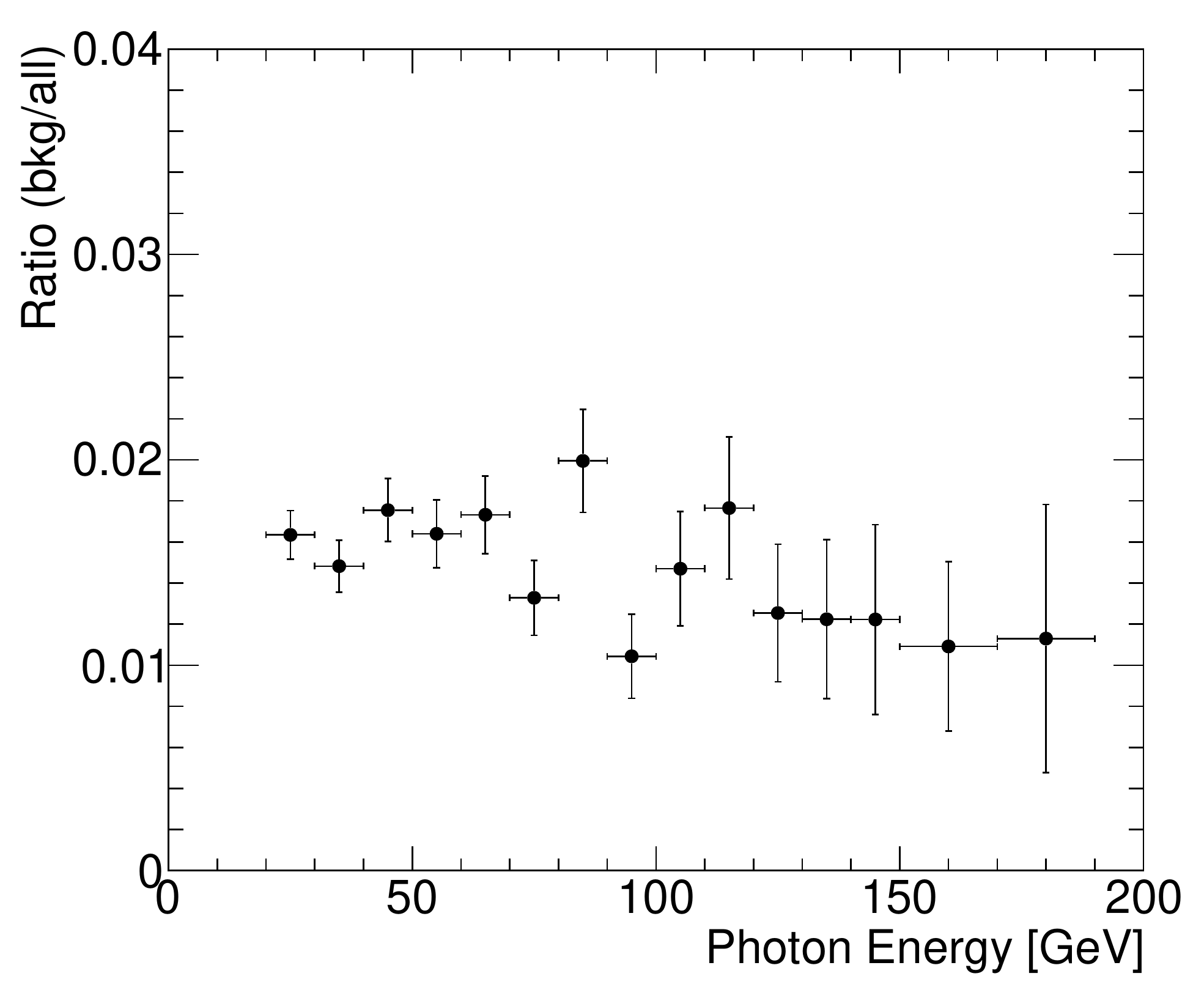}
    \caption{(Left) Bunch ID distribution of TS-shower trigger events obtained in Fill 21148. (Right) Ratio of estimated background in photon sample obtained by TS. }
    \label{fig:beamgas}
\end{figure}

The particles produced at collisions of beam particles with residual gas in the vacuum beam pipe are background in our measurements, and the background can be estimated using events associated with non-colliding bunches. Figure~\ref{fig:beamgas} represents the bunch ID distribution of the shower trigger events recorded during Fill 21148. 
The number of events at a noncolliding bunch was only ~2\% of that at a colliding bunch with the blue beam passing the IP toward the detector, which is shown as the bunch ID 31-39 in Figure ~\ref{fig:beamgas}, and almost the events resulted from the beam--gas collisions. In another set of noncolliding bunches, ID 111-119, the yellow beam passed the IP outward from the detector, and the background was negligibly small.

Considering the variation in beam intensities between the bunches, the beam--gas background in the colliding bunches, $N_{\mathrm{bkg}}$, was estimated as 
\begin{equation}
    N_{\mathrm{bkg}} = \frac{\sum_{i=(\mathrm{col.})} I_{i}}{\sum_{i=(\mathrm{non-col.})}I_{i}} N_{\mathrm{non-col.}},
\end{equation}
where $N_{\mathrm{non-col.}}$ denotes the number of events associated with the noncolliding bunches with blue beams, and $I_{i}$ represents the beam intensity for the $i$-th colliding or non-colliding bunch. 

The energy distributions of the beam--gas background events were similar to those of the signals. The right-hand side plot presented in Figure ~\ref{fig:beamgas} shows the ratio of the estimated background spectrum for photons to the photon spectrum of the colliding bunch events (signal + background) as a function of photon energy. The background fraction was approximately 1.6\% and was less influenced by the photon energy. 

\section{Performances during operation}
\label{sec:performance}
\subsection{Trigger performance}

\begin{figure}[tbp]
    \centering
    \includegraphics[width=0.9 \textwidth]{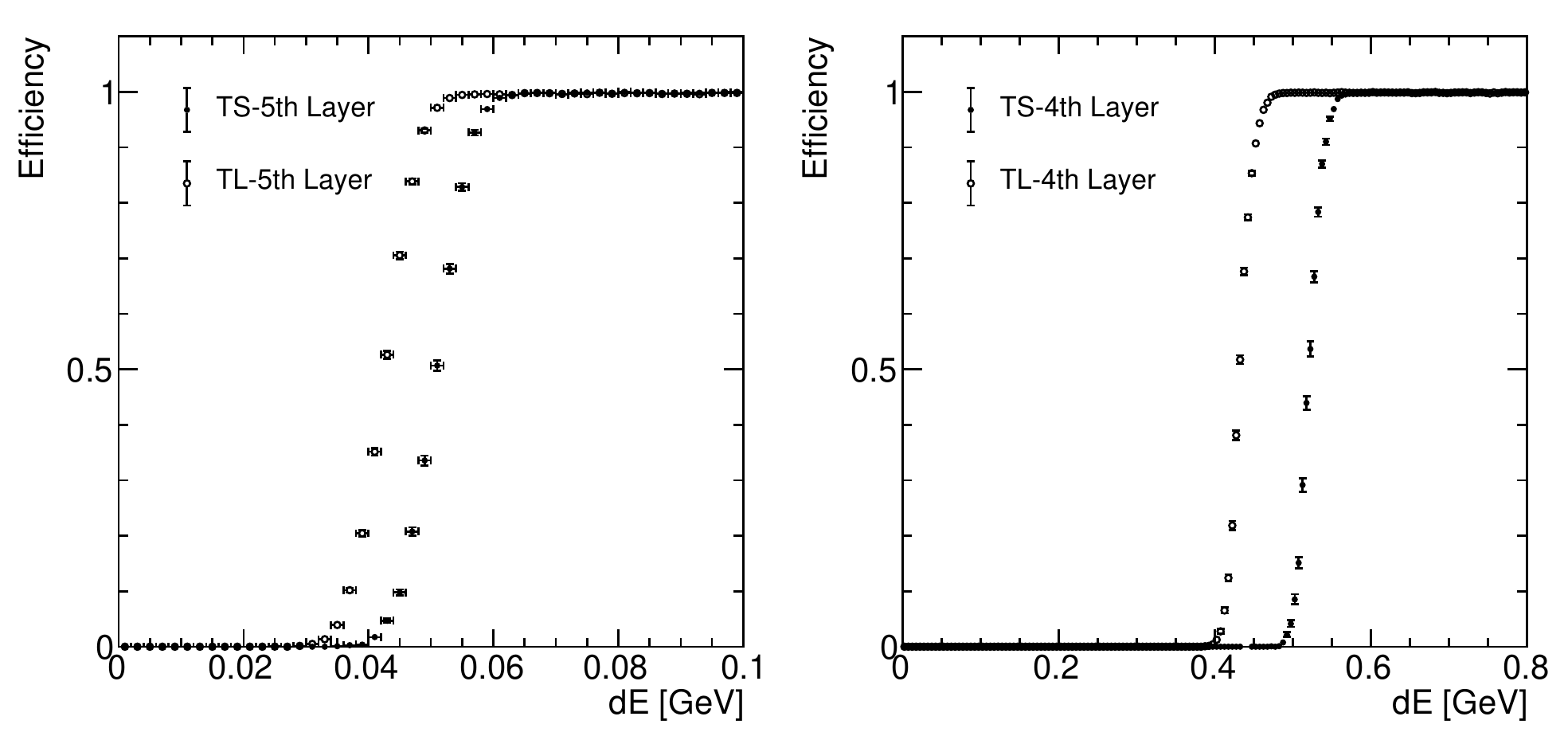}
    \caption{Threshold curves of typical discriminator channels with nominal threshold (left; fifth scintillator layer) and high threshold (right; fourth scintillator layer). Filled and open circles represent results for TS and TL, respectively.}
    \label{fig:discricurve}
\end{figure}

The efficiencies of the shower and high-EM triggers were evaluated using the obtained data and simulation. Subsequently, the triggers were generated from a combination of discriminator-hit signals, as discussed in Section ~\ref{sec:operation}, and the hit patterns of all the discriminator channels were recorded along with the calorimeter data for each event. The threshold curve of each discriminator channel was measured based on the data of the hit flag and the recorded energy deposition of the corresponding scintillator layer. The results of the fourth and fifth scintillator layers, corresponding to the high-threshold channels of ~500 MeV and the typical channels with a nominal threshold of ~45 MeV are presented in Figure ~\ref{fig:discricurve}.

\begin{figure}[tbp]
    \begin{center}
    \includegraphics[width=0.9 \textwidth]{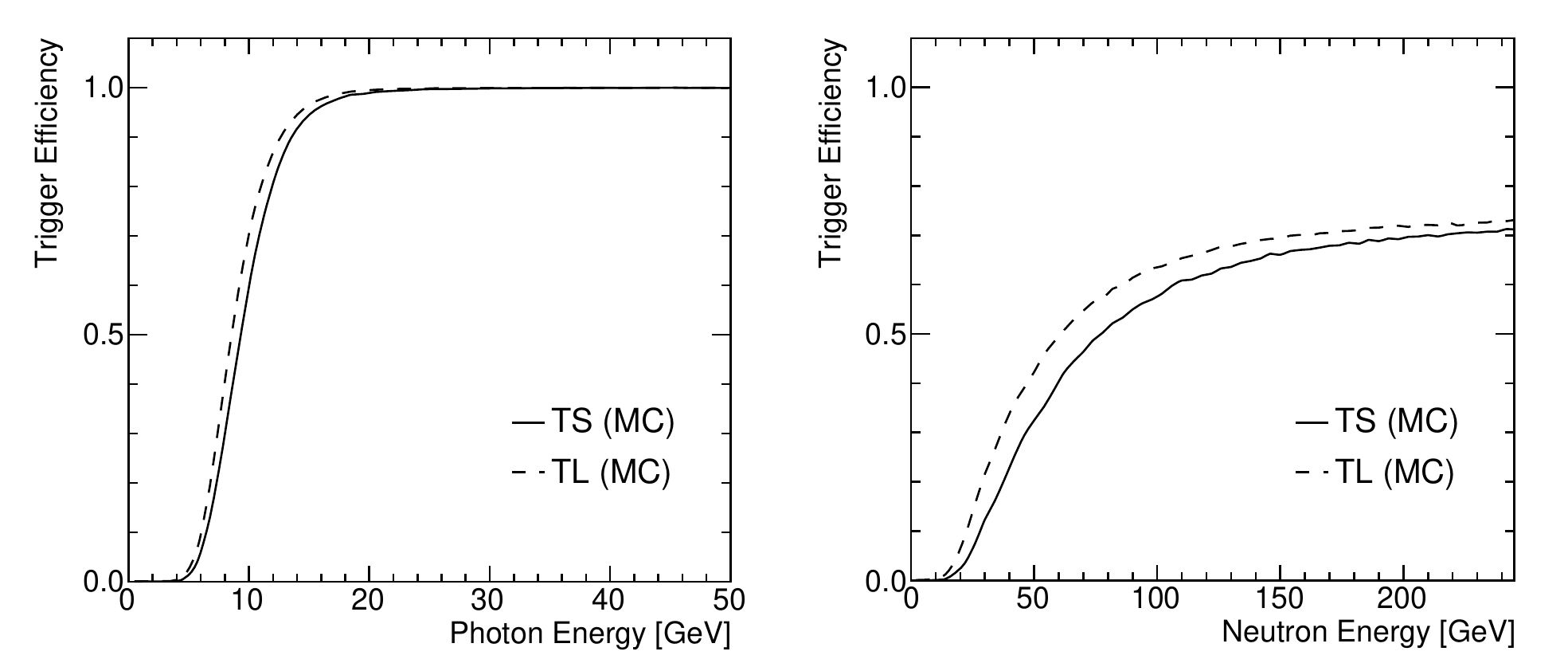} 
    \caption{ Efficiencies of shower trigger for photons (left) and neutrons (right) as a function of energy. Solid and dashed lines represent the values for TS and TL, respectively. }
    \label{fig:trigger_eff_shower}
    \end{center}
  \end{figure}

The trigger efficiencies of shower triggers for photons and neutrons were estimated by implementing the measured discriminator responses into the detector simulation, as shown in Figure ~\ref{fig:trigger_eff_shower}. 
Although the efficiency of the photons was 100\% above 20 GeV, the efficiency of neutrons gradually increased with the energy up to 100 GeV and reached approximately 70\%. The resulting inefficiency of 30\% was caused by neutrons without any interaction in the detector. 

\begin{figure}[tbp]
    \begin{center}
    \includegraphics[width=0.6 \textwidth]{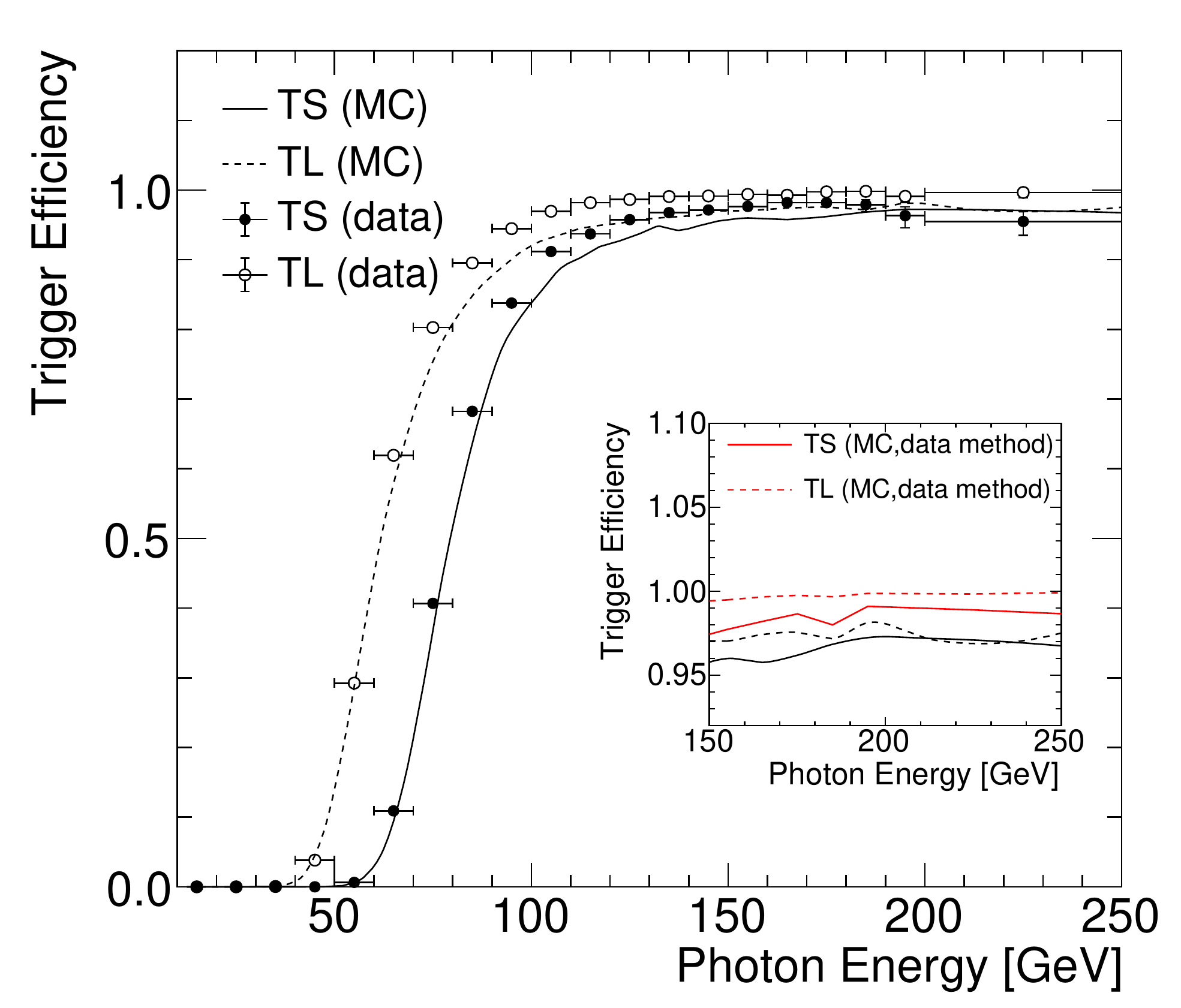} 
    \caption{Trigger efficiency of high-EM trigger for photons.
    The efficiencies were estimated using the simulation method (black solid and dashed lines) and the data method (open and filled circles). To evaluate a bias of the methods, results of the data method applying a MC sample is shown as red lines in inset. }
    \label{fig:trigger_eff_highem}
    \end{center}
  \end{figure}

In addition to the simulation method, the efficiency of the high-EM trigger was evaluated using the data as a fraction of events fulfilling the high-EM trigger condition in shower-triggered events for the discriminator hit of the fourth layer. Figure \ref{fig:trigger_eff_highem} illustrates the results of the method using photon candidate events obtained in Fill 21148 as well as the results of the MC method. 
The high-EM trigger operated to effectively collect only the high-energy photon events above 100 GeV.  
In the plateau region, the results of the data exhibit slightly higher efficiencies than the simulation results. These differences can be understood as the PID selection bias of early developing showers based on the $L_{90\%}$ criterion, and contamination of hadronic showers. The feature was confirmed by using a MC simulation (inset of Figure \ref{fig:trigger_eff_highem}).

In the plateau region, the results of the data exhibit slightly higher efficiencies than the simulation results. These differences can be understood as a PID selection bias in the data method. The inefficiency found in the simulation result originates from deeply-developed events in the shower fluctuation, while these events were excluded in the data method through the event selection based on the $L_{90\%}$ criterion discussed in Section \ref{sec:analysis}. In addition, contamination of hadronic shower events in the photon candidate sample used in the data method makes another bias. The contamination in TS, which located near the beam center, is larger than that in TL, and it induces larger inefficiency of TL than TS in the data method. The feature was confirmed by applying the data method to a detector simulation sample (inset of Figure \ref{fig:trigger_eff_highem}).

\subsection{Studies using $\pi^0$}
\label{sec:pi0}

In this section, we study the detector performance using $\pi^0$ events. A $\pi^0$ generated in a {\it pp} collision immediately decays into a photon pair with a branching ratio of 98.8\%, and the kinematics of $\pi^0$, such as the energy $E_{\pi^0}$ and the invariant mass of the system $M_{\gamma\gamma}$, which should be the $\pi^0$ rest mass of 135 MeV, can be reconstructed from the energies and hit positions of the photons measured by the detector as 
\begin{eqnarray}
    E_{\pi^0} &=& E_{\gamma_1} + E_{\gamma_2}, \\
    M_{\gamma\gamma} &\approx& \sqrt{ E_{\gamma_1} E_{\gamma_2}} \theta = \sqrt{ E_{\gamma_1} E_{\gamma_2}} \frac{D}{L}.
\end{eqnarray}
where $E_{\gamma_1}$ and $E_{\gamma_2}$ denote the energies of the photons, and $\theta$ is the opening angle between the photons, which was calculated from the distance between the hit positions on the detector, $D$, and the distance between the IP and detector, $L$. 
The $\pi^0$ candidate events were categorized into two types, called \typei and \typeii, considering the event topology. In a \typei $\pi^0$ event, each calorimeter tower observed one of the photons from a $\pi^0$ decay, and the energy was estimated using Eq.~\ref{eq:energy} with an additional correction of the shower-particle contamination leaking from one tower to the other tower. In a \typeii $\pi^0$ event, a photon pair was observed by one tower. Moreover, the total photon energy was measured using Eq.~\ref{eq:energy}, and each photon energy was determined by introducing an energy-sharing factor estimated from the lateral distributions measured by the position-sensitive layers. The \typeii $\pi^0$ event reconstruction was performed only for the larger tower (TL) owing to the small acceptance of the TS. 
Further details of the $\pi^0$ reconstruction algorithm are presented in \cite{LHCf_Pi0_Run1, RHICf_pi0_An}.

\begin{figure}[tbp]
    \begin{center}
    \includegraphics[width=0.9 \textwidth]{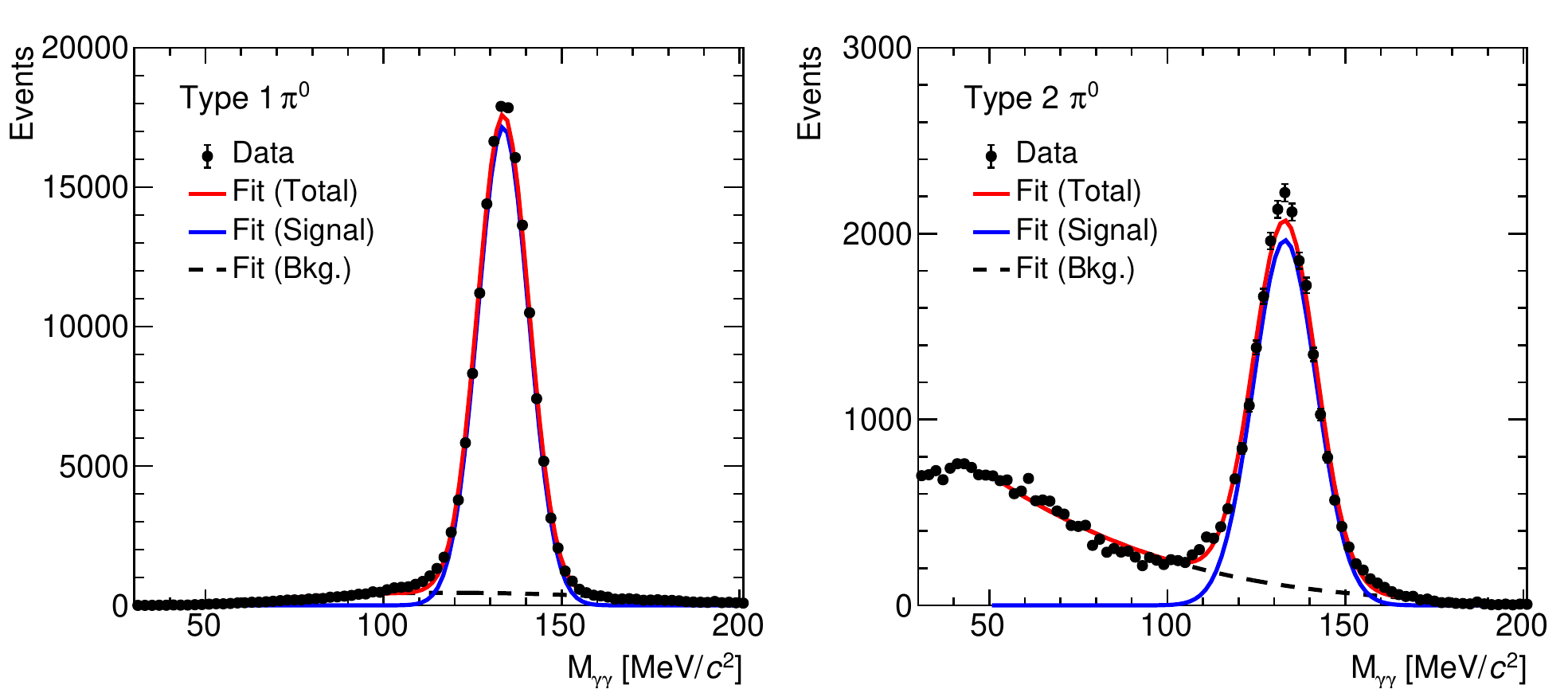} 
    \caption{Reconstructed mass distributions of \typei (left) and \typeii (right) $\pi^0$ events obtained in Fill 21148. Red lines represent the fitting result to a composite model of Gaussian and six-order polynomial functions, and these signal and background components are represented as blue and black dashed lines, respectively.} 
    \label{fig:pi0distribution}
    \end{center}
  \end{figure}

Figure \ref{fig:pi0distribution} shows the reconstructed invariant mass distributions of \typei and \typeii events. The distributions were fit to a composite model, which comprises a Gaussian function for a $\pi^0$ signal component and a six-order polynomial function for a background component to extract characteristic of the $\pi^0$ events.
The overall correction factor for the energy scale $\alpha$ in Eq.~\ref{eq:energy} was obtained by comparing the peak mass values between the data and MC. The factor shifted the energy scale by 4--6\%, and it originated from the variations in the environmental conditions such as the temperature between the operation and pre-calibrations.
The width of \typei (\typeii) $\pi^0$ mass distributions, defined as the $\sigma$ parameter of Gaussian function, was 7.2 MeV (8.9 MeV), which was larger than the MC results of 5.0 MeV (6.7 MeV). The difference results from uncertainties of the position-dependent correction factors in Equation \ref{eq:sumdE}. 
 
\begin{figure}[tbp]
    \begin{center}
    \includegraphics[width=0.9 \textwidth]{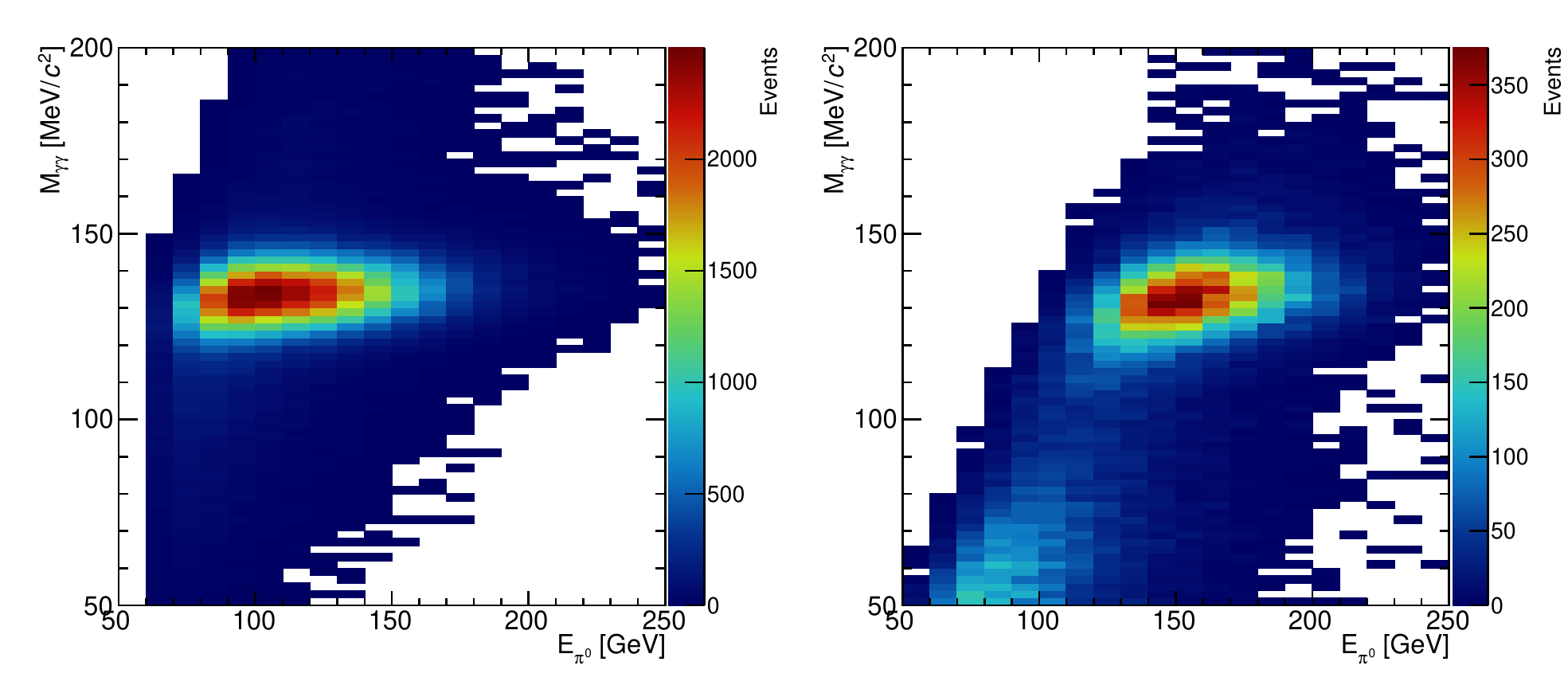} 
    \caption{ $\pi^0$ energy dependency of reconstructed mass distribution for \typei (left) and \typeii (right) $\pi^0$ events obtained in Fill 21148. } 
    \label{fig:pi0energydependency}
    \end{center}
  \end{figure}

The linearity of the energy scale was evaluated based on the dependency of the reconstructed $\pi^0$ mass on the $\pi^0$ energy.
Figure ~\ref{fig:pi0energydependency} shows $M_{\gamma\gamma}$ distributions as a function of $\pi^0$ energy. It was found that the peak-mass value of the \typeii $\pi^0$ events increased by  approximately 3\% with the $\pi^0$ energy. 
The peak-mass value of \typeii $\pi^0$ also increased by 5\%,  which was similar to 3\% increase exhibited by the MC. This implied that the stated influence originated not only from the detector response but also from the current event reconstruction algorithm, which will be improved in future. The source of the energy dependency is not understood yet, and it is considered as a systematic uncertainty in physics analyses. 

\subsection{Stability}

\begin{figure}[tbp]
    \begin{center}
    \includegraphics[width=0.9 \textwidth]{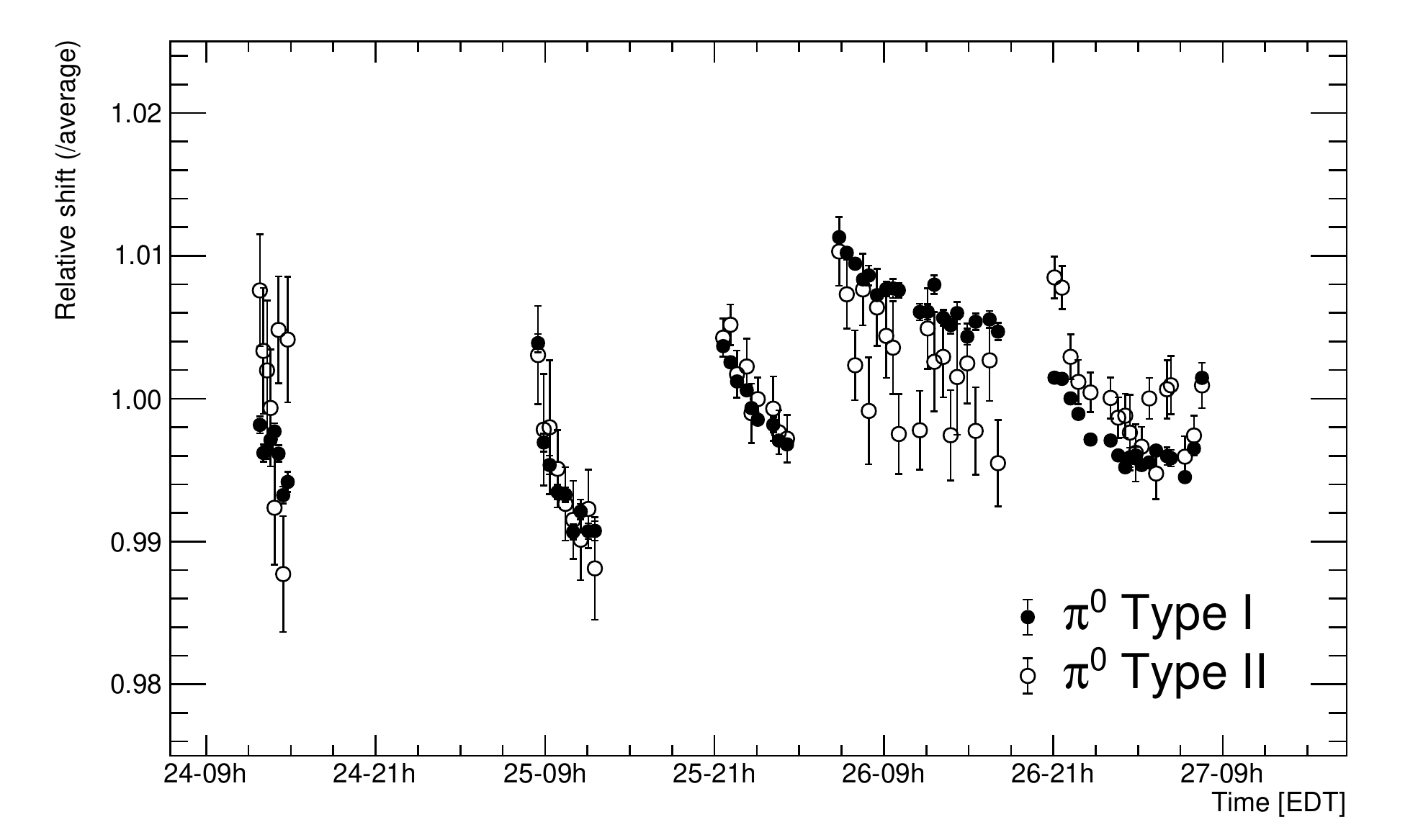} 
    \caption{Temporal variation of $\pi^0$ peak mass normalized to average value. Filled and open circles represent results of \typei and \typeii $\pi^0$ events, respectively.} 
    \label{fig:mggstability}
    \end{center}
  \end{figure}

The stability of each PMT gain during the operation was monitored using a LED light source with a precision of a few percent. The stability of the overall energy scale was evaluated more precisely based on the temporal variation of the $\pi^0$ mass peak, as shown in Figure ~\ref{fig:mggstability}. They are sufficiently stable within $\pm$1\%, whereas a small dependency on the detector position was found. The reduction of approximately 0.5\% during each fill was caused by the temperature dependency of the PMT gain with a proportional coefficient of $\sim$ -0.5 \%/degree. The temperature gradually increased after switching on the high voltages of the PMTs , which was kept off during beam injection to RHIC for safety. 

\section{Summary}
\label{sec:summary}

The RHICf operation using the LHCf-Arm1 detector at {\it pp} collisions was successfully performed with a center-of-mass collision energy of 510 GeV. During the 3-days operation, 55 M, 7.3 M, and 31 M events were obtained with the shower, $\pi^0$, and high-EM triggers, respectively.  
The performances, energy and position resolutions, and trigger efficiencies were evaluated using the obtained data and the conducted simulation, which satisfied the requirements of the physical measurements. 
In addition, the energy scale was confirmed based on $\pi^0$ events as very stable within $\pm1\%$ during the operation. 

The single spin asymmetry result of very forward $\pi^0$s was already published~\cite{RHICf_pi0_An}, and several physical analyses, production cross-section measurements for photons, $\pi^0$s, and neutrons, and single-spin asymmetry measurement for neutrons, are ongoing and will be published.
Moreover, we plan to have another operation using a newly developed detector with a silicon pad and pixel sensors to improve the performance and statistics for more precise studies of forward particle production to extend the measurement for $K^0$ to study strange meson production in the very forward region~\cite{Menjo:2021PX}. The stated operation of RHIC is scheduled in 2024.

\acknowledgments

We thank the staff at the BNL, STAR Collaboration, and PHENIX Collaboration for supporting the experiment. We especially acknowledge essential contributions of STAR members toward the successful operation of RHICf.  
This work was partly supported by the U.S.--Japan Science and Technology Cooperation Program in High-Energy Physics, JSPS KAKENHI (No. JP26247037 and No. JP18H01227), the joint research program of the Institute for Cosmic Ray Research (ICRR), University of Tokyo, and the National Research Foundation of Korea (No. 2016R1A2B2008505 and No. 2018R1A5A1025563), and “UNICT 2020-22 Linea 2” program, University of Catania.

\bibliographystyle{JHEP}
\bibliography{rhicfperformance}

\end{document}